\documentstyle[12pt]{article}
 
\begin{document}
 
\input psfig.tex
 
\begin{flushright}
DART-HEP-97/07 -- December 1997\\
\end{flushright}
 
\centerline{\bf \Large{Two Lectures on Phase Mixing}
\footnote{Invited lectures for ``Field Theoretical Tools 
in Polymer and Particle Physics'', Wuppertal, June 17-19 1997.}}
 
\vspace{0.5cm}
 
\centerline{\it Marcelo Gleiser\footnote{NSF Presidential Faculty Fellow}}
 
\vspace{ 0.5cm}
 
\centerline{Department of Physics and Astronomy}
\centerline{Dartmouth College, Hanover, NH 03755, USA
\footnote{Permanent address}}
\centerline{and}
\centerline{Nasa/Fermilab Astrophysics Center}
\centerline{Fermi National Accelerator Laboratory}
\centerline{Batavia, IL 60510}
\centerline{and}
\centerline{Osservatorio Astronomico di Roma}
\centerline{Vialle del Parco Mellini, 84, Roma I-00136 Italy}

\vskip 1.cm
\centerline{\bf Abstract}
\begin{quote}
\baselineskip 12pt

The dynamics of phase transitions plays a crucial r\^ole in the so-called
interface between high energy particle physics and cosmology. Many of the
interesting results generated during the last fifteen years or so rely
on simplified assumptions concerning the complex mechanisms typical of
nonequilibrium field theories. After reviewing well-known results
concerning the dynamics of first and second order phase transitions, I argue
that much is yet to be understood, in particular in situations where
homogeneous nucleation theory does not apply. I present a method to deal with
departures from homogeneous nucleation, and compare its efficacy with
numerical simulations. Finally, I discuss the interesting problem of matching
numerical simulations of stochastic field theories with continuum models.
\end{quote}

\def\beq{\begin{equation}}
\def\eeq{\end{equation}}
\def\ba{\begin{eqnarray}}
\def\ea{\end{eqnarray}}
\def\re#1{[{\ref{#1}]}}
 
\def\mpl{{m_{Pl}}}
\def\x{{\bf x}}
\def\p{\phi}
\def\F{\Phi}
\def\s{\sigma}
\def\a{\alpha}
\def\d{\delta}
\def\t{\tau}
\def\r{\rho}
\def\n{n(R,t)} 

\vfill\eject

\centerline{\bf LECTURE I}
\vspace{.5cm}
\section{Homogeneous Nucleation}

The fact that the gauge symmetries describing particle interactions
can be restored at high enough temperatures has led, during the past
15 years or so, to an active research program on the possible
implications that this symmetry restoration might have had to the
physics of the very early Universe. One of the most interesting and
popular possibilities is that during its expansion the Universe
underwent a series of phase transitions, as some higher symmetry group
was successively broken into products of smaller groups, up to the
present standard model described by the product $SU(3)_C\otimes
SU(2)_L \otimes U(1)_Y$. Most models of inflation and the formation of
topological (and nontopological) defects are well-known consequences
of taking the existence of cosmological phase transitions seriously
\cite{KT}.
 
One, but certainly not the only,
motivation of the works  addressed in this talk
comes from the possibility
that the baryon asymmetry of the Universe could have been dynamically
generated during a first order electroweak phase transition \cite{EW}.
As is by now clear, a realistic calculation of the net baryon number
produced during the transition is a formidable challenge. We probably
must invoke physics beyond the standard model (an exciting prospect
for most people), push perturbation theory to its limits
(and beyond, due to the nonperturbative nature of magnetic plasma
masses that regulate the perturbative expansion in the symmetric
phase), and we must deal with nonequilibrium aspects of the phase
transition. Here I will focus on the latter problem, as it seems to
me to be the least discussed of the pillars on which most baryon
number calculations are built upon. To be more specific, it is possible
to separate the nonequilibrium aspects of the phase transition into two
main subdivisions. If the transition proceeds by bubble nucleation, we
can study the propagation of bubbles in the hot plasma and the
transport properties of particles through the bubble wall. A
considerable amount of work has been devoted to this issue, and the
reader can consult the works of Ref. \cite{BW} for details. These
works assume that homogeneous nucleation theory is adequate to
investigate the evolution of the phase transition, at least for the
range of parameters of interest in the particular model being used to
generate the baryon asymmetry. This brings us to the second important
aspect of the nonequilibrium dynamics of first order phase
transitions, namely the validity of homogeneous nucleation theory to
describe the approach to equilibrium. This is the issue addressed in
this talk.

Nucleation theory is a well-studied, but far from exhausted, subject.
Since the pioneering work of Becker and D\"oring on the nucleation of
droplets in supercooled vapor \cite{BD}, the study of first order
phase transitions has been of interest to investigators in several
fields, from meteorology and materials science to quantum field theory
and cosmology. Phenomenological field theories were developed by Cahn
and Hilliard and by Langer \cite{CH,LANGER} in the context of
coarse-grained time-dependent Ginzburg-Landau models, in which an
expression for the decay rate per unit volume was obtained by assuming
a steady-state probability current flowing through the saddle-point of
the free-energy functional \cite{LANGER,DOMB}. The application of metastable
decay to quantum field theory was initiated by Voloshin, Kobzarev, and
Okun \cite{VKO}, and soon after put onto firmer theoretical ground by
Coleman and Callan \cite{CC}. The generalization of these results for
finite temperature field theory was first studied by Linde
\cite{LINDE}, and has been the focus of much recent attention
\cite{FINITETDECAY}.

The crucial ingredient in the evaluation of the decay rate is the
computation of the imaginary part of the free energy. As shown by
Langer \cite{LANGER}, the decay rate ${\cal R}$ is proportional to the
imaginary part of the free energy ${\cal F}$,
\beq
{\cal R} = {{\mid E_-\mid }\over {\pi T}}{\rm Im} {\cal F} ~,
\eeq
where $E_-$ is the negative eigenvalue related to metastability, which
depends on nonequilibrium aspects of the dynamics, such as the
growth rate of the critical bubble.
Since ${\cal F}= - T {\rm ln}
Z$, where $Z$ is the partition function, the computation for the rate
boils down to the evaluation of the partition function for the system
comprised of critical bubbles of the lower energy phase inside the metastable
phase.

If we imagine the space of all
possible field
configurations for a given model,
there will be different paths to go from the metastable
to the ground state. We can think of the two states as being separated
by a hill of a given ``height''.
The energy barrier for the decay is then related
to the height of this hill. At the top of the hill, only one direction
leads down to the ground state, the unstable direction. Fluctuations
about this direction will grow, with rate given by the negative
eigenvalue which appears in the above formula. All other directions are
positively curved, and fluctuations about them give rise to positive
eigenvalues which do not contribute to the decay rate.
The path which will cost less energy is the one which will
dominate the partition function, the so-called critical bubble or bounce.
It is simply the field configuration that interpolates between the two
stable points in the {\it energy landscape}, the metastable and ground state.
The energy barrier for the decay is the energy of this particular field
configuration.
 
For a dilute gas of bubbles only, the partition function for
several bubbles is given by \cite{AM,LANGER},
\ba
Z & \simeq & Z(\varphi_{f}) +Z(\varphi_{f}) \left[ \frac{Z(\varphi_{b})}
{Z(\varphi_{f})} \right] + Z( \varphi_{f}) \frac{1}{2 !} \left[
\frac{Z(\varphi_{b})}{Z(\varphi_{f})} \right]^{2} + \ldots
\nonumber \\
& \simeq & Z(\varphi_{f}) \exp \left[ \frac{Z(\varphi_{b})}{Z(\varphi_{f})}
\right]
\: ,
\label{e:Zmany}
\ea
where $\varphi_{f}$ is the metastable vacuum field configuration and
$\varphi_{b}$ is the bubble configuration, the bounce solution to the
$O(3)$-symmetric Euclidean equation of motion. We must evaluate the
partition functions above. This is done by the saddle-point method,
expanding the scalar field $\phi({\bf x},\tau)$,
such that $\phi({\bf x},\tau) \rightarrow \varphi_{f}
+\zeta(\bf{x} ,\tau)$ for $Z(\varphi_{f})$, and $ \phi({\bf x},\tau)
\rightarrow \varphi_{b} (\bf{x}) +\eta (\bf{x} ,\tau)$ for
$Z(\varphi_{b})$, where $\zeta(\bf{x} ,\tau)$ and $\eta (\bf{x} ,\tau)$
are small fluctuations about equilibrium.

It is crucial to note that the saddle-point, or Gaussian, method only
gives good results if indeed the fluctuations about
equilibrium are sufficiently small that nonlinear terms in the fields
can be neglected. Even though the method sums over all amplitude fluctuations,
it does so by assuming that the functional integral
is well approximated by truncating
the expansion of the action to second order. The efficiency of the method
relies on the fact that higher amplitudes will be suppressed fast enough
that their contribution to the partition function will be negligible. One
can visualize this by comparing a sharp parabolic curve with a flatter
one with minimum at $x_0$,
and investigating when $\int dx e^{-f(x)}$ will be well approximated
by writing $f(x)\simeq f(x_0) + {1\over 2}(x-x_0)^2f^{\prime\prime}(x_0)$.
For a sharp curve, larger amplitude fluctuations will be strongly
suppressed and thus give a negligible contribution to
the integral over all amplitudes. Clearly, this will not be the case for
flatter curves.
 
Skipping details \cite{FINITETDECAY}, using the saddle-point method
one obtains for the ratio of partition functions,
$\frac{Z(\varphi_{b})}{Z(\varphi_{f})}$,
\begin{equation}
\frac{Z(\varphi_{b})}{Z(\varphi_{f})} \stackrel{saddle-point}{\simeq}
\left[ \frac{\det ( -\Box_{E} + V''(\varphi_{b}))_{\beta}}
{ \det ( -\Box_{E} + V''(\varphi_{f}))_{\beta}} \right]^{-\frac{1}{2}}
e^{-\Delta S} \: ,
\label{ratiodet}
\end{equation}
where $[ \det (M)_{\beta}]^{- \frac{1}{2}} \equiv \int D \eta \exp
\left\{ - \int_{0}^{\beta} d \tau \int d^{3} x \frac{1}{2} \eta [M]
\eta \right\}$ and $\Delta S = S_{E}(\varphi_{b})-S_{E}(\varphi_{f})$
is the difference between the Euclidean actions for the field
configurations $\varphi_{b}$ and $\varphi_{f}$. [Note that
$S_{E}(\varphi)$, and hence $\Delta S$, does not include any
temperature corrections. It would if one had summed over other fields
coupled to $\varphi$.] Thus, the free energy of the system is,
\beq\label{energy}
{\cal F} = - T \left[ \frac{ \det ( -\Box_{E} + V''(\varphi_{b}))_{\beta}}
{ \det ( -\Box_{E} + V''(\varphi_{f}))_{\beta}} \right]^{-\frac{1}{2}}
e^{-\Delta S} \: .
\eeq

Let me stress again the assumptions that go into computing the free energy.
First, that
the partition function is given by Eq. \ref{e:Zmany} within the
dilute gas approximation, and second, that the partition function is
evaluated approximately by
assuming {\it small} fluctuations about the homogeneous
metastable state $\varphi_{f}$. It is clear that for situations in
which there are large amplitude fluctuations about the metastable
equilibrium state the above formula must break down. Thus the
breakdown of the expression for the rate is intimately connected with
the question of how well-localized the system is about the metastable
state as the temperature drops below the critical temperature $T_c$.
Homogeneous nucleation, as its name already states, is only accurate when
the metastable state is sufficiently homogeneous. In the presence of
inhomogeneities, there is no reason to expect that the decay rate formula
will apply. The question then is to quantify {\it when} does it
break down and how can we incorporate nonperturbative corrections
to the decay rate in a consistent way.

\section{Nonperturbative corrections to decay rates}

In order to investigate the importance of large-amplitude fluctuations in the
description of first order phase transitions, I have developed numerical
simulations in two \cite{MG}
and, with J. Borrill, three \cite{BG}
spatial dimensions, which measured the fraction 
of the volume of the system in the initial phase as a function of the barrier
height. Since these have been documented elsewhere, here I quickly describe
the main idea and results. 

Imagine a scalar field with a degenerate double-well potential. The field
is coupled to a thermal bath at temperature $T$ through a Langevin-like
equation which assumes that the bath is Markovian, {\it i.e.}, the noise
is white and additive. The system is artificially divided into two regions,
left and right of the maximum of the potential, call it the negative and
positive regions, respectively. The system is prepared initially in one of the
regions, say, the negative region with $\phi({\bf x}, t=0)=-1$. 
The coupling to the bath will then drive
fluctuations of the field around this minimum and we measure the fraction
of the total volume in each of the two regions as a function of the parameters
controlling the height of the potential barrier, usually the temperature
and/or a coupling constant.

We observed that while for large enough potential barriers the system remained
well-localized around its initial state, a sharp change of behavior
occurred for a critical value of the specific
control parameter being varied. In the
case examined in Ref. \cite{BG}, the control parameter was the quartic
coupling of the scalar field, $\lambda$. We showed that for $\lambda >
\lambda_c$ the system became completely mixed, in that the volume was
equally shared by the positive and negative regions. In other words,
for $\lambda > \lambda_c$,
the effective potential describing the system is not a degenerate 
double-well, but a parabolic curve centered at $\langle \phi \rangle=0$;
Thermal fluctuations have ``restored'' the symmetry of the system. 

The challenge was thus to model the large amplitude fluctuations which
were responsible for this phase mixing. In what follows I review the 
so-called subcritical bubbles method which can model {\it quantitatively}
the dynamics of large, nonperturbative, thermal fluctuations in scalar
field theories.

\vfill\eject

\centerline{\bf LECTURE II}
\vspace{.5cm}
\subsection{Modeling nonperturbative fluctuations: Symmetry restoration
and phase mixing}

As was stressed before, the computation of decay rates based on homogeneous
nucleation theory assumes a smooth metastable background over which critical
bubbles of the lower free energy phase will appear, grow and coalesce,
as the phase transition
evolves. However, as the results from the numerical simulations indicate,
the assumption of smoothness is not always valid. To the skeptical reader,
I point out that several condensed matter experiments indicate
that homogeneous nucleation fails to describe the transition
when the nucleation barrier ($\Delta S/T$) becomes too small. Furthermore,
the agreement between theory and experiment has a long and problematic
history \cite{NUCEXP}. Homogeneous nucleation has to be used with care, in
a case by case basis. 

The basic idea is that in a hot system, not only small but also large
amplitude fluctuations away from equilibrium will, in principle,
be present. Small
amplitude fluctuations are perturbatively
incorporated in the evaluation of the finite
temperature effective potential, following well-known procedures. Large
amplitude fluctuations probing the nonlinearities of the theory are not.
Whenever they are important, the perturbative effective potential becomes
unreliable. In an ideal world, we should be able to sum over all amplitude
fluctuations to obtain the exact partition function of the model, and
thus compute the thermodynamic quantities of interest. However, we can only
to this perturbatively, and will always miss information coming from the
fluctuations not included in its evaluation. If large amplitude fluctuations
are strongly suppressed, they will not contribute to the partition function,
and we are in good shape. But what if they are important, as argued above?
We can try to
approach this question avoiding complicated issues related to the
evaluation of path integrals beyond the Gaussian approximation by obtaining
a kinetic equation which describes the fraction of volume populated by these
large amplitude fluctuations. In order to keep the treatment simple, and
thus easy to apply,
several assumptions are made along the way, which I believe are quite sensible.
In any case, the strength of the method is demonstrated when the results are
compared with the numerical experiments described before.

Large amplitude fluctuations away from equilibrium are modelled by
Gaussian-profile spherically-symmetric
field configurations of a given size and amplitude. They can be thought of
as being coreless bubbles. Keeping
with the notation of the numerical experiment,  fluctuations away from the
0-phase [called the ``negative phase'' above], and into the 0-phase 
are written respectively as,
 
\beq
\phi_c(r)=\phi_ce^{-r^2/R^2}~,~~~\phi_0(r)=\phi_c\left (1-
e^{-r^2/R^2}\right ) \:,
\eeq
where $R$ is the radial size of the configuration, and $\phi_c$ is the
value of the amplitude at the bubble's core,
away from the 0-phase. In previous treatments
(cf. Refs. \cite{GKW} and \cite{GG}), it was assumed that $\phi_c=\phi_+$,
that is, that the configuration interpolated between the two minima of the
effective potential, and that $R=\xi(T)$, where $\xi(T)=m(T)^{-1}$
is the
mean-field correlation length. But in general, one should sum over all
radii and amplitudes above given
values which depend on the particular model under
study. This will become clear as we go along.

Define $dn(R,\phi,t)$
as the number density of bubbles of radius between $R$ and
$R+dR$ at time $t$, with amplitudes $\phi\geq \phi_c$ between $\phi$ and
$\phi+d\phi$. By choosing to sum over bubbles of amplitudes $\phi_c$
and larger, we are effectively describing the system as a ``two-phase''
system. For example, in the numerical simulation above it was assumed
that the negative-phase was for amplitudes $\phi \leq \phi_{\rm max}$, and the
positive-phase was for amplitudes $\phi > \phi_{\rm max}$. Clearly, for a
continuous system this division is artificial. However, since the models
we are interested in have two local minima of the free energy, this
division becomes better justified. Fluctuations with small enough amplitude
about the minima
are already summed over in the computation of the effective potential.
It is the large amplitude ones which are of relevance here. To simplify the
notation, from now on I will denote by ``+ phase''
all fluctuations with amplitudes
$\phi > \phi_c$ and larger. The choice of $\phi_c$ is model-dependent, as will
be clear when we apply this formalism to specific examples.

The fact that the bubbles shrink will be incorporated in the
time dependence for the radius $R$\footnote{Of course, the amplitude $\phi$
will also be time-dependent. However, its time-dependence is coupled to
that of the radius, as recent studies have shown \cite{OSC}.
In order to
describe the effect of shrinking on the
population of bubbles it is sufficient to include only the
time dependence of the radius.}.
Here, I will
only describe a somewhat simplified approach to the dynamics. More details
are provided in the work by Gleiser, Heckler, and Kolb \cite{GHK}.
The results, however, are essentially identical.

The net rate at which bubbles of a given
radius and amplitude are created and destroyed is given by the kinetic
equation,
\begin{eqnarray}
\label{eq:KIN}
{{\partial \n}\over {\partial t}}=-{{\partial \n}\over {\partial R}}
\left ({{dR}\over {dt}}\right )+\left ({{V_0}\over {V_T}}\right )\Gamma_{0
\rightarrow +}(R)  \nonumber\\
 - \left ({{V_+}\over {V_T}}\right )\Gamma_{+\rightarrow 0}(R)
\end{eqnarray}
Here, $\Gamma_{0\rightarrow +}(R)$ ($\Gamma_{+\rightarrow 0}(R)$)
is the rate per unit volume for
the thermal nucleation of a bubble of radius $R$ of positive-phase  within
the 0-phase (0-phase within
the positive-phase). 
$V_{0(+)}$ is the volume available for nucleating bubbles of the
+(0) phase. Thus we can write, for the total volume of the system,
$V_T=V_0+V_+$, expressing the fact that the system has been ``divided'' into
two available phases, related to the local minima of the free energy density.
It is convenient to define the fraction of volume in the + phase,
$\gamma$, as

\beq
{{V_+}\over {V_T}} \equiv \gamma = 1-{{V_0}\over {V_T}}~~.
\eeq
 
In order to compute $\gamma$ we must sum over {\it all} bubbles of different
sizes, shapes, and amplitudes within the + phase, {\it i.e.}, starting with
$\p_{\rm min} \geq \p_c$.
Clearly, we cannot compute $\gamma$ exactly. But it turns out
that a very good approximation is obtained by assuming that the bubbles are
spherically symmetric, and with radii above a given minimum radius, $R_{\rm
min}$. The reason we claim that the approximation is good comes from comparing
the results of this analytical approach with numerical simulations.
The approximation starts to break down as the
background becomes more and more mixed, and the morphology of the ``bubbles''
becomes increasingly
more important, as well as other terms in the kinetic equation which
were ignored. For example, there should be a term which accounts for
bubble coalescence, which increases the value of $\gamma$. This term becomes
important when the density of bubbles is high enough for the probability
of two or more of them coalescing to be non-negligible. As we will see,
by this point the mixing is already so pronounced that we are justified in
neglecting this additional complication to the kinetic equation. As a bonus,
we will be able to solve it analytically. The expression for $\gamma$ is,
 
\beq\label{e:gamma}
\gamma \simeq \int_{\p_{\rm min}}^{\infty}\int_{R_{\rm min}}^{\infty}
\left ({{4\pi R^3}\over 3}\right ){{\partial^2 n}\over {\partial \p\partial R}}
d\p dR~~.
\eeq
 
The attentive reader must have by now noticed that we have a coupled
system of equations; $\gamma$, which appears in the rate equation for the
number density $n$, depends on $n$ itself. And, to make things even worse,
they both depend on time. Approximations are in order,
if we want to proceed any further along an analytical approach. The first thing
to do is to look for the equilibrium solutions, obtained by setting
$\partial n / \partial t =0$ in the kinetic equation. In equilibrium, $\gamma$
will also be constant\footnote{This doesn't mean that thermal
activity in or between the two
phases is frozen; equilibrium is a statement of the average
distribution of thermodynamical quantities. Locally, bubbles will be
created and destroyed, but always in such a way that the average value of
$n$ and $\gamma$ are constant.}. If wished, after finding the equilibrium
solutions one can find the time-dependent solutions, as was done in
Ref. \cite{GG}. Here, we are only interested in the final equilibrium
distribution of subcritical bubbles, as opposed to the approach to
equilibrium.
 
The first approximation is to take the
shrinking velocity of the bubbles to be constant, $dR/dt = -v$. This is in
general not the case ({\it cf.} Ref. \cite{OSC}),
but it does encompass the fact that subcritical bubbles
shrink into oblivion.
The strength of the thermodynamic approach is that details of how
the bubbles disappear are unimportant, only the time-scale playing a r\^ole.
The second approximation is to assume that the rates for creation and
destruction of subcritical fluctuations are Boltzmann suppressed, so that
we can write them as $\Gamma= AT^4e^{-F_{\rm sc}/T}$, where $A$ is an arbitrary
constant of order unity, and $F_{\rm sc}(R,\p_c)$
is the cost in free energy to
produce a configuration of given radius $R$ and core amplitude $\p_c$. For the
Gaussian {\it ansatz} we are using, $F_{\rm sc}$ assumes the general
form, $F_{\rm sc}=\alpha R +\beta R^3$, where $\alpha = b\p_c^2$ ($b$ is
a combination of $\pi$'s and other numerical factors)
and $\beta$ depends on
the particular potential used. In practice, the cubic term can usually be
neglected, as the free energy of small ($R\sim \xi$) subcritical bubbles
is dominated by the gradient (linear) term.
We chose to look at the system at the
critical temperature $T_c$. For this temperature, the creation and destruction
rates, $\Gamma_{0\rightarrow +}$ and $\Gamma_{+\rightarrow 0}$ are identical.
Also, for $T_c$, the approximation of neglecting the cubic term is very good
(in fact it is better and better the larger the bubble is)
even for large bubbles, since for degenerate vacua
there is no gain (or loss) of volume energy for
large bubbles. Finally, we use that $V_+/V_T = \gamma$ in the $\Gamma_{+
\rightarrow 0}$ term.
A more sophisticated
approach is presented in Ref. \cite{GHK}.

We can then write the equilibrium rate equation as,
 
\beq
{{\partial n}\over {\partial R}} =  - cf(R)~~,
\eeq
where,
 
\beq
c\equiv (1-2\gamma)AT^4/v,~~f(R)\equiv e^{-F_{\rm sc}/T}~~.
\eeq
 
Integrating from $R_{\rm min}$ and imposing that $n(R\rightarrow\infty)=0$,
the solution is easily found to be,
 
\beq
n(R) = {c\over {\alpha(\p_c)/T }}e^{-\alpha(\p_c) R/T}~~.
\eeq
 
Not surprisingly, the equilibrium number density of bubbles is Boltzmann
suppressed. But we now must go back to $\gamma$, which is buried in the
definition of $c$. We can solve for $\gamma$ perturbatively, by plugging
the solution for $n$ back into Eq. \ref{e:gamma}. After a couple of
fairly nasty integrals, we obtain,
 
\beq\label{e:gamma_sol}
\gamma = {{g\left (\alpha(\p_{\rm min}),R_{\rm min}\right ))}\over
{1 + 2 g\left (\alpha(\p_{\rm min}),
R_{\rm min}\right )}}~~,
\eeq
where,
 
\beq
g\left (\alpha(\p_{\rm min}),R_{\rm min}\right )
 = {{4\pi}\over 3}\left ({{AT^4}\over v}\right )
\left ({T\over {\alpha}}\right )^3{{e^{-\alpha R_{\rm min}/T}}\over {
\alpha/T}}\left [6+\left ({{\alpha R_{\rm min}}\over T}\right )^3
+3{{\alpha R_{\rm min}}\over T}\left (2+{{\alpha R_{\rm min}}\over T}\right )
\right ]~~.
\eeq

We can now apply this formalism to any model we wish. The first obvious
application is to compare $\gamma$ obtained from the numerical experiment
with the value obtained from the kinetic approach. From the definition
of the equilibrium fractional population difference, 
$\Delta F_{\rm EQ}\left (\theta_c \right ) \equiv f_0^{\rm eq} - f_+^{\rm eq}$,
 
\beq
\Delta F_{\rm EQ}\left (\theta_c \right ) = 1 -2\gamma~~.
\eeq
 
Thus, it is straightforward to extract the value of $\gamma$
from the numerical simulations as a function of $\lambda$.
Also, as we neglected the volume contribution to the free energy of
subcritical bubbles, we have,
 
\beq
F_{sc} = \alpha(\p_c)R_{\rm min} = {{3\sqrt{2}}\over 8}\pi^{3/2}X_-^2
(\theta_c)R_{\rm min} ~~,
\eeq
where, as you recall, $X_-$ is the position of the maximum of the mean-field
potential used in the simulations. So, we must sum over all amplitudes
with $X\geq X_-$, and all radii with $ R\geq 1$ (in dimensionless units),
as we took the
lattice spacing to be $\ell=1$. That is,
we sum over all possible sizes, down to
the minimum cut-off size of the lattice used in the simulations. In practice,
we simply substitute $\p_c=X_-$ and $R_{\rm min}=1$ in the expression
for
$\gamma$.
In Fig. 1, we compare the numerical
results for $\gamma$ (dots) with the results from the analytical integration
of the kinetic equation. The plots are for different values
of the parameter $A/v$.
Up to the critical
value for $\lambda\simeq 0.025$, 
the agreement is very convincing. As we increase
$\lambda$ into the mixed phase region of the diagram, the kinetic approach
underestimates the amount of volume in the +-phase. This is not surprising,
since for these values of $\lambda$
the density of subcritical bubbles is high enough that
terms not included in the equation become important, as I mentioned before.
However, the lack of agreement for higher values of $\lambda$ is irrelevant,
if we are interested in having a measure of the
smoothness of the background; clearly, the rise in $\gamma$ is sharp enough
that homogeneous nucleation should not be trusted for $\lambda > 0.024$ or so,
as the fraction of volume occupied by the +-phase is already around 30\%
of the total volume. Subcritical bubbles give a simple and quantitatively
accurate picture
of the degree of inhomogeneity of the background, offering a guideline as to
when homogeneous nucleation theory can be applied with confidence, or,
alternatively, when the effective potential needs higher order corrections.

\subsection{Modeling nonperturbative fluctuations: ``Inhomogeneous'' nucleation}

Now we apply the subcritical bubbles method to the decay of metastable states
in the case that the homogeneous nucleation formalism (section I) does not
apply. Details of this work can be found in Ref. \cite{GH}.

As mentioned before,
if there is significant phase mixing in the background metastable state,
its free-energy density is no longer $V(\phi=0)$, where I assume the
potential has a metastable state at $\phi=0$.
One must also account for the free-energy density of the nonperturbative,
large-amplitude fluctuations. Since there is no formal way of deriving
this contribution outside improved perturbative schemes, we
propose to estimate the corrections to the background free-energy
density by following another route. We start by writing the free energy
density of the metastable state as
$V(\phi=0) +
{\cal F}_{\rm sc}$,
where ${\cal F}_{\rm sc}$ is the nonperturbative contribution to
the free-energy density due to the large amplitude fluctuations, which we
assume can be modelled by
subcritical bubbles. We will calculate  ${\cal F}_{\rm sc}$
further below.

\begin{figure}
\hspace{1.in}
\psfig{figure=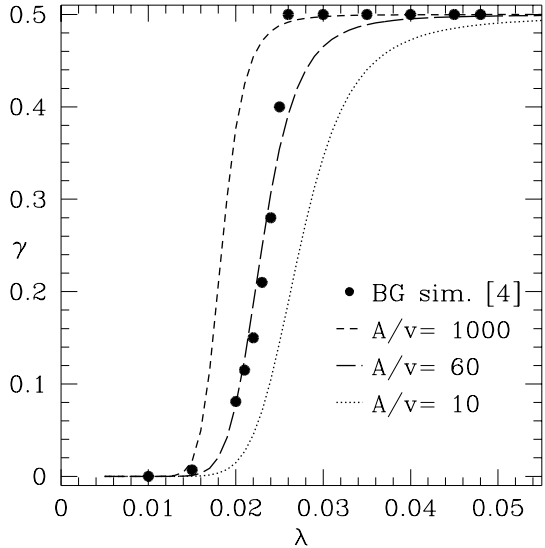,width=3.in,height=3.in}
\caption{The fraction of the volume in the + phase.
The dots are from numerical simulations of 
Ref. 14, while
the lines are the solutions of the Boltzmann eq. for different values
of the parameter $A/|v|$.}
\end{figure}

We thus define the effective free-energy difference between the two minima,
$\Delta V_{\rm cg}$, which includes
corrections due to phase mixing, as
\begin{eqnarray}\label{DeltaV}
\Delta V_{\rm cg}= \Delta V_{0} + {\cal F}_{\rm sc}
\end{eqnarray}
which is the sum of the free-energy difference calculated in
the standard way,
and the ``extra'' free-energy density
due to the presence of subcritical bubbles. Henceforth, the subscript `cg'
will stand for ``coarse-grained''.

Since for degenerate
potentials (temperature-dependent or not)
no critical bubbles should be nucleated, taking into
account subcritical bubbles must lead to a change in the coarse-grained
free-energy density (or potential)
describing the transition. Thus, it should be possible to translate the
``extra'' free energy available in the system due to the presence of
subcritical bubbles in the background into a corrected potential for the
scalar order parameter. We will write this corrected potential as
$ V_{\rm cg}(\phi)$.

The standard coarse-grained free energy is calculated by integrating out
the short-wavelength modes (usually up to the correlation length)
from the partition function
of the system, and is approximated by
the familiar form \cite{Langer74a}
\begin{eqnarray}\label{F}
F_{\rm cg}(\phi) = \int d^3r \left( \frac{1}{2}(\nabla \phi)^2 +
V_{\rm cg}(\phi)\right)~.
\end{eqnarray}
How do we estimate $V_{\rm cg}$? One way
is to simply constrain it to be consistent with the thin wall limit.
That is, as $V_{\rm cg}(\phi)$ approaches degeneracy
({\it i.e.} $\Delta V_{\rm cg}(\phi)\rightarrow 0$), it must obey the
thin wall limit of eq.~(\ref{DeltaV}). Note that with a simple rescaling,
a general polynomial potential (to fourth order)
can be written in terms of one free parameter.
Thus, the thin
wall constraint can be used to express the corrected value of
this parameter in terms of ${\cal F}_{\rm sc}$ in appropriate units.
The free energy of the critical bubble is then obtained
by finding the bounce solution to the equation of motion
$\nabla^{2}\phi - dV_{\rm cg}(\phi)/d\phi=0$ by the usual shooting method,
and substituting this solution
into  eq.~(\ref{F}).

In order to determine $V_{\rm cg}$, we must first calculate the
free-energy density ${\cal F}_{\rm sc}$ of the subcritical bubbles. From the
formalism presented in the previous subsection,
\begin{eqnarray}\label{calF}
{\cal F}_{\rm sc} \approx  \int_{\phi_{\rm min}}^{\infty}
{\int_{R_{\rm min}}^{R_{\rm max}}{F_{\rm sb}
\frac{\partial^{2}\n}{\partial R \partial \phi_{A}}dRd\phi_{A}}},
\end{eqnarray}
where $\phi_{\rm min}$ defines the lowest amplitude within the +phase,
typically (but not necessarilly)
taken to be the maximum of the double-well potential. $R_{\rm min}$ is the
smallest radius for the subcritical bubbles, compatible with the
coarse-graining scale. For example, it can be
a lattice cut-off in numerical simulations,
or the mean-field correlation length in continuum models. As for $R_{\rm max}$,
it is natural to choose it to be the critical bubble radius.

\begin{figure}
\hspace{1.in}
\psfig{figure=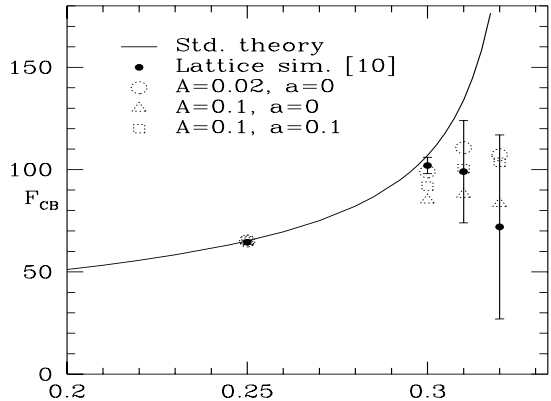,width=3.in,height=2.3in}
\caption{Comparison between numerical data and theoretical
predictions for the decay barrier with and without the
inclusion of subcritical bubbles. The parameter $a$ is related to an
extra term in the Boltzmann eq. which can be safely neglected.}
\end{figure}

As an application of the above method, we investigated nucleation rates
in the context of a 2-d model for which accurate numerical results
are available \cite{Alford93a}.
This allowed us to compare the results obtained by incorporating
subcritical bubbles into the calculation of the decay barrier with the
results from the numerical simulations. The potential used was
written in terms of one dimensionless
parameter $\lambda\equiv
m^2h/g^2$,
\begin{eqnarray}\label{V2d}
V(\phi) = \frac{1}{2} \phi^2 - \frac{1}{6} \phi^{3 }
+\frac{\lambda}{24} \phi^4.
\end{eqnarray}
This double-well potential is degenerate when $\lambda = 1/3$,
and the second minimum is lower than the first when $\lambda <1/3$.

As argued before,
we find the new coarse-grained potential $V_{\rm cg}$
(or, equivalently, $\lambda_{\rm cg}$) by constraining it to agree with the
thin wall limit. Simple algebra from eqs.~(\ref{DeltaV}) and (\ref{V2d})
yields, to first order in the deviation from degeneracy,
\begin{eqnarray}\label{}
\lambda_{\rm cg} = \lambda - \frac{{\tilde {\cal F}}_{\rm sc}}{54}
\end{eqnarray}
where ${\tilde {\cal F}}_{\rm sc}=\frac{g^2}{m^6}{\cal F}_{\rm sc}$ is
the dimensionless free-energy density in subcritical bubbles.
The new potential $V_{\rm cg}$ is then used to
find the bounce solution and the free energy of the critical bubble.

In Fig. 2 we show that the calculation of the nucleation barrier including
the effects of subcritical bubbles is consistent with data from lattice
simulations,
whereas the standard calculation overestimates the barrier by a large margin.
In fact, the inclusion of subcritical bubbles provides a reasonable
explanation for the anomalously high nucleation rates observed in the
simulations close to degeneracy.

\section{Matching numerical simulations to continuum field theories}

As a final topic to be discussed in this lecture, I would like to change gears
and briefly turn to the issue of how to match numerical simulations of field
theories with their continuum counterparts. In particular, I am interested in
situations where a degree of stochasticity is present in the simulations,
as for example happens when we model the coupling of fields to a thermal
or quantum bath via a Langevin-like equation, or even in the form of noisy
initial conditions.

Although field theories are continuous
and usually formulated in an infinite volume, lattice simulations are discrete
and finite, imposing both a maximum (``size of the box'' $L$) and a minimum
(lattice spacing $\delta x$) wavelength that can be probed by the simulation.
When the  system is coupled to an external thermal (or quantum) bath,
fluctuations will be constrained within the allowed window of wavelengths,
leading to discrepancies between the continuum formulation of the theory and
its lattice simulations; the results will be dependent on the
choice of lattice spacing.

Parisi suggested that if proper
counterterms were used, this depedence on lattice spacing could be
attenuated \cite{PARISI}.
Recently, Borrill and Gleiser (BG)
have examined this question within the context of 2-d critical
phenomena \cite{BGII}.
They have computed the counterterms needed to render the
simulations indepedent of lattice spacing and have obtained a match between
the simulations and the continuum field theory, valid within the one-loop
approximation used in their approach. Here, I want to focus mostly on the
application of these techniques to 1-d field theories, in 
particular to the problem of thermal nucleation of kink-antikink pairs.
[This is based on work of Ref. \cite{GM}.]

Even though 1-d field theories are
free of ultra-violet divergences, the
ultra-violet cutoff imposed by the lattice spacing will generate a
{\it finite} contribution to the effective potential which must
be taken into account if we are
to obtain a proper match between the theory
and its numerical simulation on a
discrete lattice. If neglected, this contribution may compromise the
measurement of physical quantities such as the density of kink-antikink pairs
or the effective kink mass.

For classical, 1-dimensional finite-temperature field theories,
the one-loop corrected effective potential
is given by the momentum integral \cite{PARISI}
\begin{equation}\label{e.oneloopdef}
V_{\rm 1L}(\phi) = V_0(\phi) + {T\over 2}\int_0^{\infty}{{dk}\over
{2\pi}}{\rm ln}\left [1 + {V_0''(\phi)\over k^2} \right ]
= V_0(\phi) + {T\over 4}\sqrt{V_0''(\phi)} \; .
\end{equation}

As mentioned before, the lattice spacing $\delta x$
and the lattice size $L$
introduce long and short momentum cutoffs
$\Lambda = \pi / \delta x$ and $k_{\rm min} = 2 \pi / L$, respectively.
Lattice simulations are characterized by one dimensionless parameter,
the number of degrees of freedom $N = L/\delta x$.
For sufficiently large $L$ one can neglect the effect of $k_{\rm min}$
and integrate from $0$ to $\Lambda$.
For $V_0''\ll\Lambda^2$ (satisfied for sufficiently large $\Lambda$),
the result can be expanded into
\begin{equation}\label{e.1Lseries}
V_{\rm 1L}(\phi,\Lambda) = V_0 + {T\over 4}\sqrt{V_0''} -
{T\over 4\pi}  {V_0''\over \Lambda} +
\Lambda T\,\, {\cal O}\left( V_0''^2\over \Lambda^4 \right) \, .
\end{equation}

As is to be expected for a 1-dimensional system,
the limit $\Lambda \rightarrow \infty$ exists and is well-behaved;
there is no need for renormalization of ultra-violet divergences.
However, the effective one-loop potential
is lattice-spacing dependent through the explicit appearance
of $\Lambda$, and so are the corresponding numerical
simulations.
In order to remove this dependence on $\delta x$, we
follow the renormalization procedure given by
BG \cite{BGII};
it is irrelevant if the
$\Lambda$-dependent terms are ultra-violet
finite ($d=1$) or infinite ($d\geq 2$).
In the lattice formulation of the theory, we
add a (finite) counterterm to the tree-level potential
$V_0$ to  remove the lattice-spacing dependence of the results,
\begin{equation}\label{e.Vct.gen}
V_{\rm ct}(\phi)=
{T\over 4\pi}  {V_0''(\phi)\over \Lambda}\; .
\end{equation}

There is an additional, $\Lambda$-independent, counterterm which
was set to zero by an appropriate choice of renormalization scale.
The lattice simulation then uses the corrected potential
\begin{equation}\label{e.Vlatt.gen}
V_{\rm latt}(\phi)=V_0(\phi) +
{T \delta x\over 4\pi^2} V_0''(\phi)\; .
\end{equation}

Note that the above treatment yields two novel results. First, that
the use of $V_{\rm latt}$ instead of $V_0$ gets rid of the dependence
of simulations on lattice spacing. [Of course, as $\delta x\rightarrow 0$,
$V_{\rm latt}\rightarrow V_0$. However,
this limit is often not computationally efficient.] 
Second, that the effective interactions
that are simulated must be compared to
the one-loop corrected potential
$V_{\rm 1L}(\phi)$ of Eq.\ (\ref{e.oneloopdef}); once the
lattice formulation is made independent of lattice
spacing by the addition of the proper
counterterm(s), it simulates,
within its domain of validity, the thermally corrected
one-loop effective potential.

\begin{figure}
\begin{minipage}{.46\linewidth}
\psfig{figure=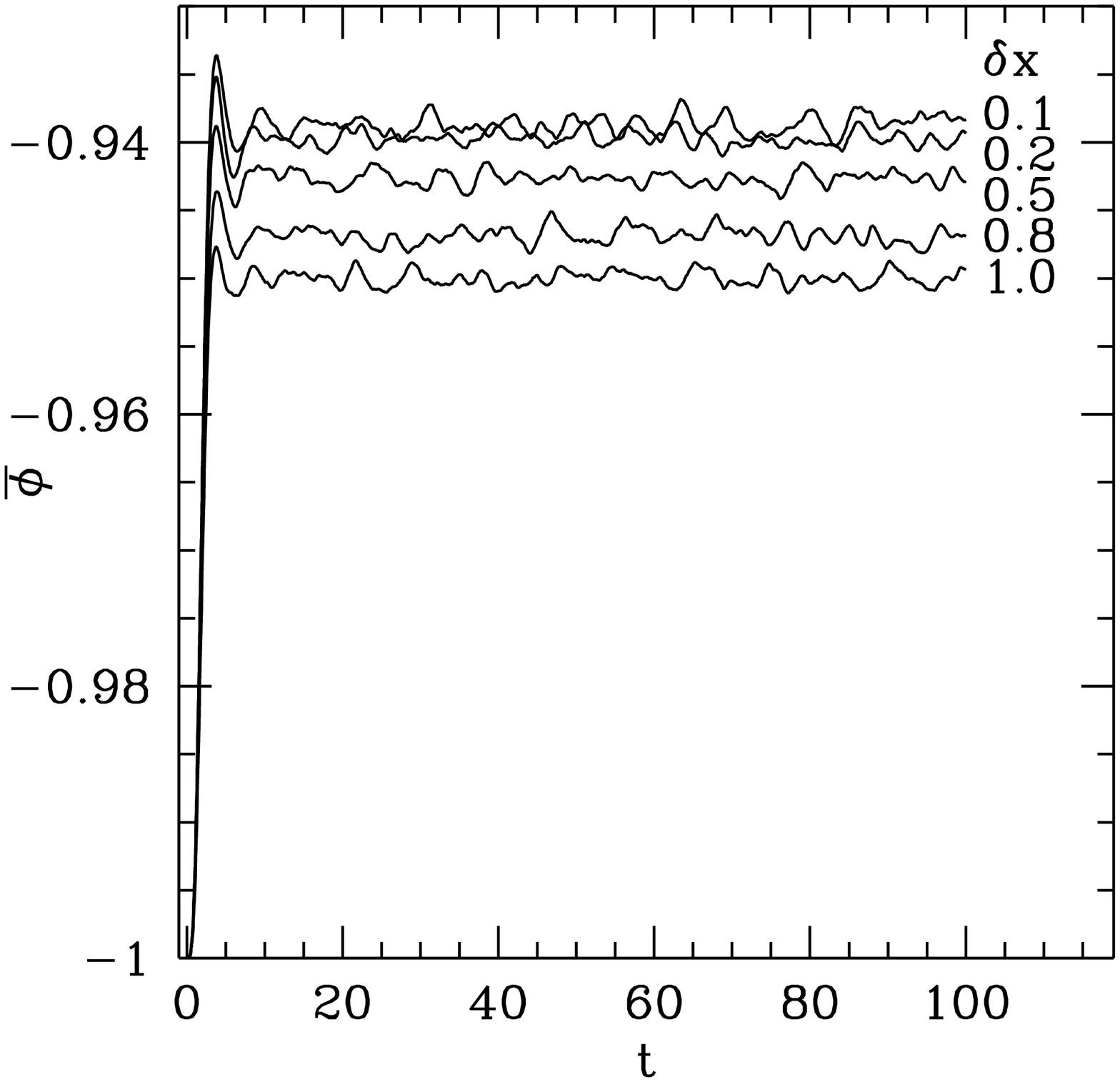,width=\linewidth}
\end{minipage}
\begin{minipage}{.46\linewidth}
\psfig{figure=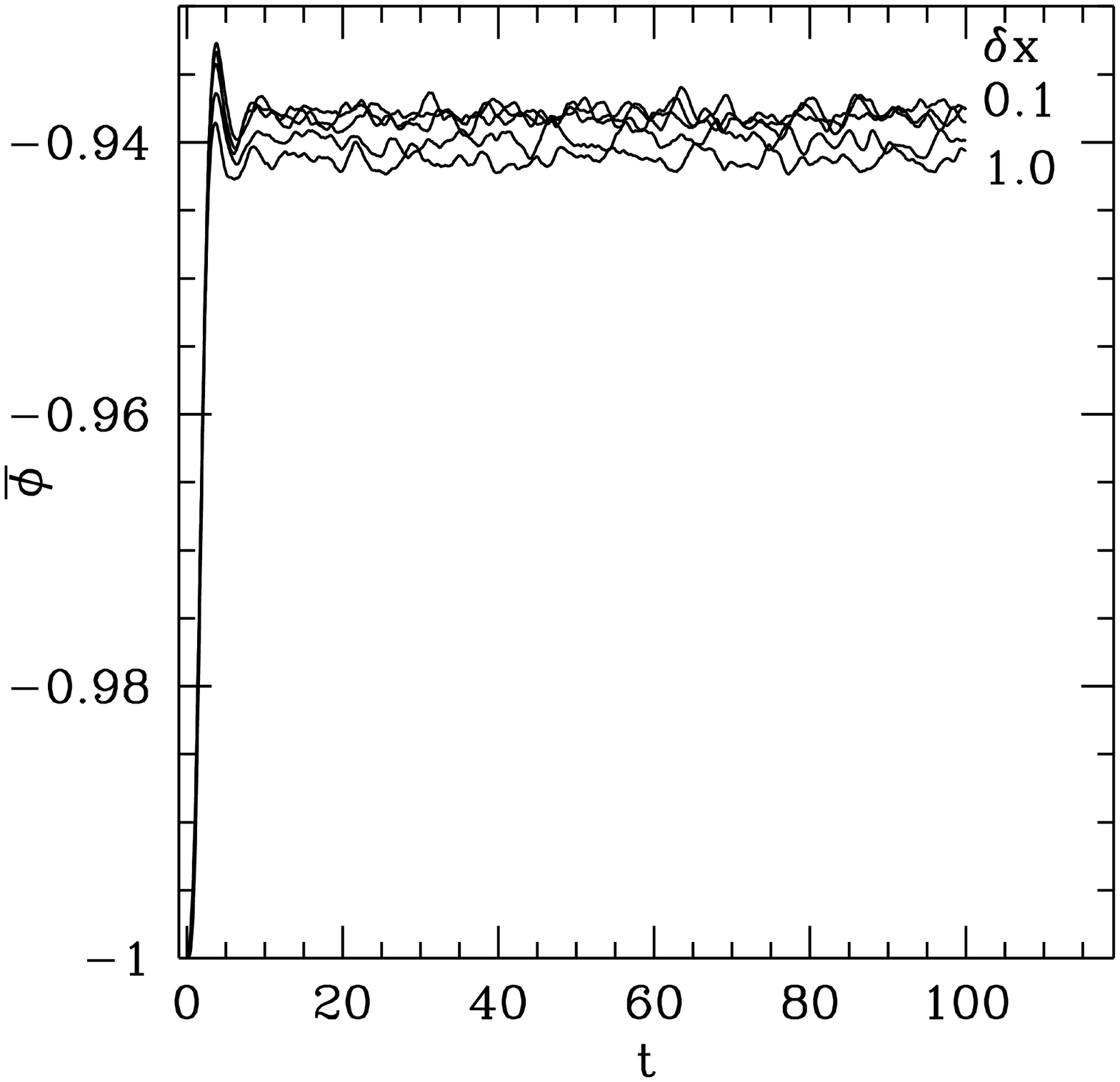,width=\linewidth}
\end{minipage}
\caption{Average field value $\bar\phi (t)$ for $T=0.1$
using the tree-level potential,
left, and the corrected potential, right. The filter cutoff is $\Delta L=3$.
\label{f.phiave}}
\end{figure}

Applying this method to the formation of kink-antikink pairs, we get a 
corrected potential,
\begin{equation}\label{e.VLatt}
V_{\rm latt}(\phi) =
V_0(\phi) + {3\over 4 \pi^2} \lambda T \delta x \phi^2\; ;
\end{equation}
simulations using $V_{\rm latt}$ will, in principle, match the continuum theory
\begin{equation}\label{e.V1L}
V_{\rm 1L}(\phi) =
V_0(\phi) + {T \sqrt{\lambda}\over 4} \sqrt{3\phi^2 - \phi_0^2}\; ,
\end{equation}
which has (shifted) minima at
$\pm\phi_{\rm min}(T)$, with $\phi_{\rm min}(T) < \phi_0$.

As a first test of our procedure, we investigate the mean field
value $\bar\phi (t)=(1/L) \int\phi (x,t) dx$ {\it before} the nucleation of a
kink-antikink pair, {\it i.e.}, while the field is still well localized in
the bottom of the well.
In Fig. 3 we show the ensemble average of $\bar\phi$
(after 100 experiments) for different values of $\delta x$, ranging
from 1 down to 0.1, at $T = 0.1$.
The simulations leading to the left graphs use the ``bare'' potential
$V_0$, whereas the right graphs are produced employing $V_{\rm latt}$
(Eq.\ \ref{e.VLatt}).

Perhaps the most difficult task when counting the number of kink-antikink
pairs that emerge during a simulation is the identification of what precisely is
a kink-antikink pair at
different temperatures. Typically, we can identify
three ``types'' of fluctuations: i) small amplitude, perturbative fluctuations
about one of the two minima of the potential; ii) full-blown kink-antikink
pairs interpolating between the two minima of the potential;
iii) nonperturbative fluctuations which have large amplitude but not quite
large enough to
satisfy the boundary conditions required for a kink-antikink pair. These latter
fluctuations are usually dealt with by a smearing of the field over a certain
length scale. Basically, one chooses a given smearing length $\Delta L$ which
will be large enough to ``iron out'' these ``undesirable'' fluctuations but not
too large that actual kink-antikink pairs are also ironed-out. The choice of
$\Delta L$ is, in a sense, more an art than a science, given our ignorance
of how to handle these nonperturbative fluctuations.

The smearing was implemented as a low pass filter with filtering cutoff
$\Delta L$; the field is Fourier transformed, filtered at a given
wavelength, and Fourier 
transformed back.
We counted pairs by identifying the zeros of the
filtered field. 
Choosing the filter cutoff
length to be too large may actually undercount the number of pairs.
Choosing it too low may include nonpertubative fluctuations as pairs. We
chose $\Delta L=3$ in the present work, as this is the smallest ``size'' for
a kink-antikink pair. In contrast, in the works by Alexander et al. a
different method was adopted, that looked for zero-crossings for eight lattice
units (they used $\delta x=0.5$) to the left and right of a zero crossing
\cite{HABIB}.
We have checked that our simulations reproduce the results of Alexander et al.
if we: i) use the bare potential in the lattice simulations and ii) use a
large filter cutoff length $\Delta L$. Specifically, the number of pairs found
with the bare potential for $T=0.2,~\delta x=0.5$ are: $n_p = 36,~ 30,~{\rm
and}~ 27$, for $\Delta L = 3,~5,~{\rm and}~ 7$ respectively. Alexander et al.
found (for our lattice length) $n_p = 25$. Comparing results for different
$\Delta L$,
it is clear that the differences between our results and those of Alexander
et al. come from using a different potential in the simulations, {\it viz.}
a dressed vs. a bare potential. For small $\delta x$ these differences
disappear.

\begin{figure}
\begin{minipage}{.46\linewidth}
\psfig{figure=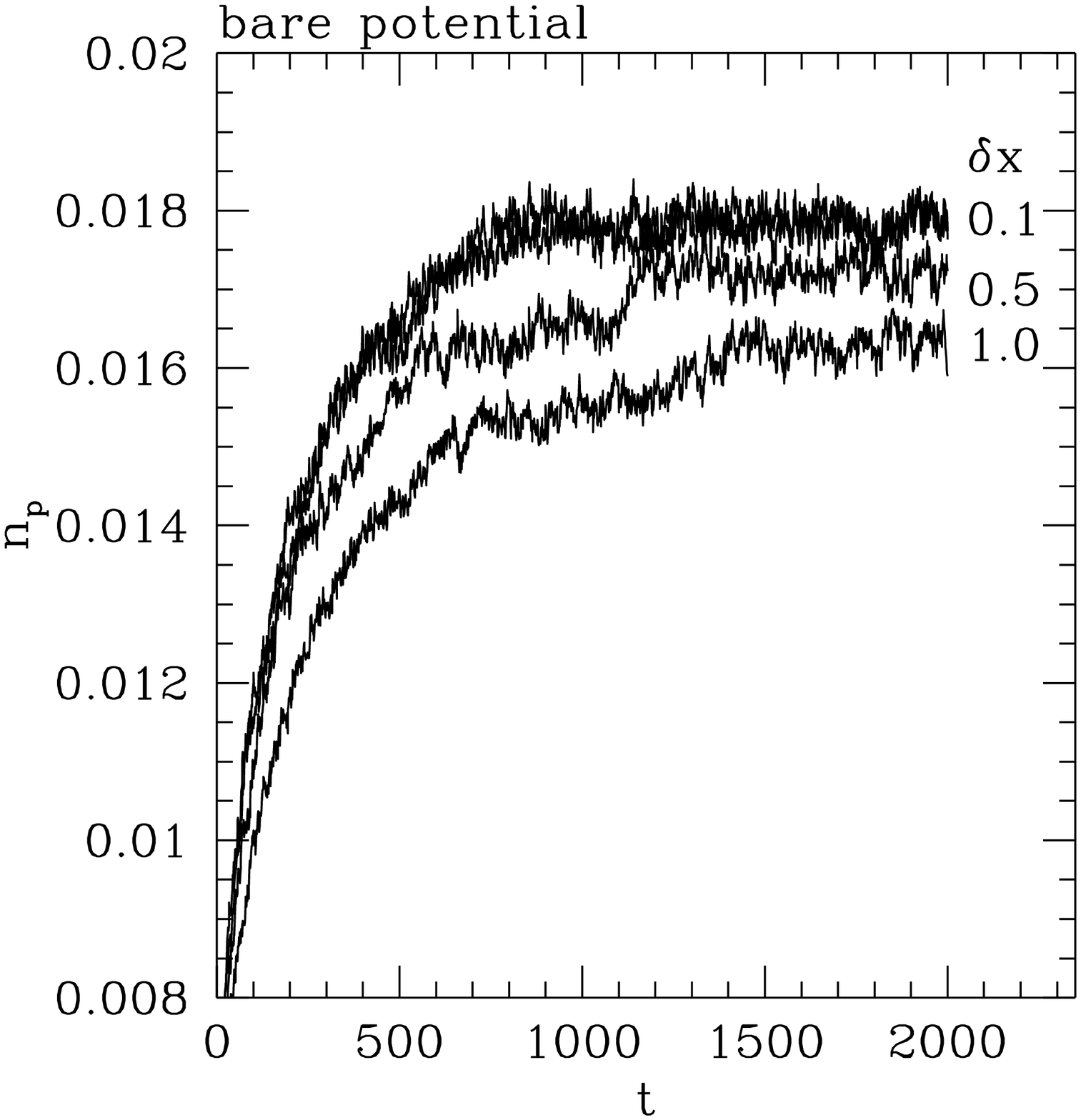,width=\linewidth}
\end{minipage}
\begin{minipage}{.46\linewidth}
\psfig{figure=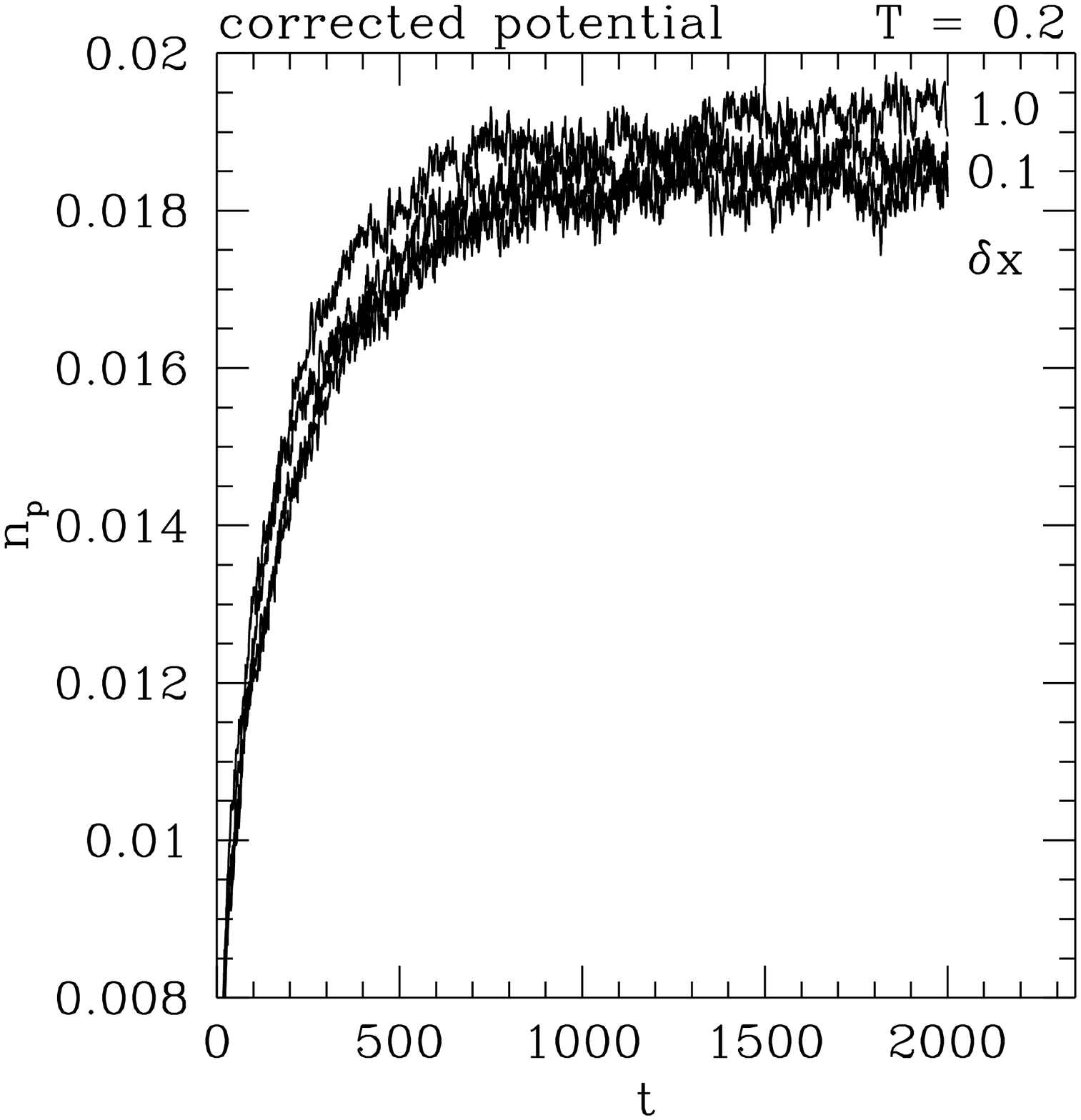,width=\linewidth}
\end{minipage}
\caption{\label{f.kinkdens}Density of kink-antikinks
(half of density of zeros), for $T=0.2$
and $\delta x = 1$, 0.5, 0.2, and 0.1. The filter cutoff is $\Delta L=3$.}
\end{figure}

Fig. 4 compares measurements of the kink-antikink
pair density (half
the number of zeros of the smeared field), ensemble-averaged
over 100 experiments,
for different lattice spacings.
Again it is clear from the graphs on the left
that using the tree-level potential
$V_0$ in the
simulations causes the results to be dependent on $\delta x$,
whereas the addition of the finite counterterm
removes this problem quite efficiently.

Another step is to establish what is the continuum theory being simulated. Due
to space limitations, I refer the reader to the work of Ref. \cite{GM} for
more details.

This work was written in part while the author was visiting the Osservatorio 
Astronomico di Roma and the Nasa/Fermilab
Astrophysics Center. I thank Franco Occhionero and Luca Amendola 
for their kind hospitality in Rome and Josh Frieman et al. for their
hospitality at Fermilab.
The author was partially supported by the
National Science Foundation through a
Presidential Faculty Fellows Award no. PHY-9453431 and by the
National Aeronautics and Space Administration through
grant no. NAGW-4270.


\begin{thebibliography}{}


\bibitem{KT} 
E. W. Kolb and M. S. Turner (1990):
{\it The Early Universe}
(Addison-Wesley, New York, 1990).

\bibitem{EW}
A. G. Cohen, D. B. Kaplan, and A. E. Nelson, {\it
Annu. Rev. Nucl. Part.
Sci.} {\bf 43}, 27 (1993); 
{A. Dolgov}, {\it Phys. Rep.} {\bf 222}, 311 (1992).

\bibitem{BW}  
B. Liu, L. McLerran, and N. Turok, {\it Phys. Rev.}
{\bf D46}, 2668 (1992); A. G. Cohen, D. B. Kaplan, and A. E. Nelson, {\it
Phys. Lett.} {\bf B336}, 41 (1994);
M. B. Gavela,
P. Hern\'andez, J. Orloff, and O. P\`ene, {\it Mod. Phys. Lett.} {\bf A9},
795 (1994);
P. Huet and E. Sather, {\it Phys. Rev.} {\bf D51}, 379 (1994).

\bibitem{BD}
R. Becker and W. D\"oring, {\it Ann. Phys.} {\bf 24}, 719 (1935).

\bibitem{CH}
J. W. Cahn and J. E. Hilliard,
             {\sl J. Chem. Phys.} {\bf 31}, 688 (1959).

\bibitem{LANGER}  
J. S. Langer, {\it Ann. Phys. (NY)} {\bf 41}, 108 (1967);
{\it ibid.} {\bf 54}, 258 (1969).

\bibitem{DOMB}
J. D. Gunton, M. San Miguel and P. S. Sahni, in
{\it Phase Transitions and
Critical Phenomena}, {\bf Vol. 8}, Ed. C. Domb and J. L. Lebowitz (Academic
Press, London, 1983).

\bibitem{VKO}
M. B. Voloshin, I. Yu. Kobzarev, and L. B. Okun',
        {\sl Yad. Fiz.} {\bf 20}, 1229 (1974)
        [Sov. J. Nucl. Phys. {\bf 20}, 644 (1975) ].

\bibitem{CC}
S. Coleman, {\it Phys. Rev.} {\bf D15}, 2929 (1977);
C. Callan and S. Coleman, {\it Phys. Rev.} {\bf D16}, 1762 (1977).

\bibitem{LINDE}
A. D. Linde,  {\it Phys. Lett.} {\bf 70B}, 306 (1977);
{\it Nucl. Phys.} {\bf B216}, 421 (1983);
[Erratum: {\bf
B223}, 544 (1983)].

\bibitem{FINITETDECAY}
M. Gleiser, G. Marques, and R. Ramos, {\it Phys. Rev.}
{\bf D48}, 1571 (1993); D. E. Brahm and C. Lee, {\it Phys. Rev.}
 {\bf D49}, 4094 (1994); D. Boyanovsky, D. E. Brahm, R. Holman, and
D.-S. Lee, {\it Nucl. Phys.} {\bf B441}, 609 (1995).

\bibitem{AM} 
P. Arnold and L. McLerran, {\it Phys. Rev. }{\bf D36}, 581
(1987); {\it ibid.} {\bf D37}, 1020 (1988).



\bibitem{MG}
M. Gleiser, {\it Phys. Rev. Lett.} {\bf 73}, 3495 (1994).

\bibitem{BG}
J. Borrill and M. Gleiser, {\it Phys. Rev.} {\bf D51},
4111 (1995).

\bibitem{NUCEXP}
E.D. Siebert and C.M. Knobler, {\it Phys. Rev. Lett.},
{\bf 52}, 1133 (1984); J.S. Langer and A.J. Schwartz, {\it Phys. Rev.}
{\bf A21}, 948 (1980); A. Leggett in {\it Helium Three}, ed. by
W.P. Halperin and L.P. Pitaevskii, (North-Holland, New York, 1990);
for an (outdated) review of the situation in the early eighties see
Ref. \cite{DOMB}.

\bibitem{GKW}
M. Gleiser, E. W. Kolb, and R. Watkins, {\it Nucl. Phys.}
{\bf B364}, 411 (1991); G. Gelmini and M. Gleiser,
{\it Nucl. Phys.} {\bf B419}, 129 (1994);
M. Gleiser and E. W. Kolb,
 Phys. Rev. Lett. {\bf 69}, 1304 (1992); N. Tetradis, {\it Z. Phys.}
{\bf C57}, 331 (1993).

\bibitem{GG}
G. Gelmini and M. Gleiser in Ref. \cite{GKW}.

\bibitem{OSC}
M. Gleiser, Phys. Rev. {\bf D49}, 2978 (1994); E.J. Copeland, M. Gleiser,
and H.-R. M\"uller, Phys. Rev. {\bf D52}, 1920 (1995).

\bibitem{GHK}
M. Gleiser, A. Heckler, and E.W. Kolb, Phys. Lett. B405
(1997) 121.

\bibitem{Langer74a}
J.S. Langer, Physica {\bf 73}, 61 (1974).


\bibitem{GH}
A. Heckler and M. Gleiser, Phys. Rev. Lett. {\bf 76},
180--183 (1996).

\bibitem{Alford93a}
M. Alford and M. Gleiser, Phys. Rev. {\bf D48},
2838--2844 (1993).

\bibitem{PARISI} 
G. Parisi, {\it Statistical Field Theory} (Addison-Wesley,
New York, 1988).

\bibitem{BGII}
J. Borrill and M. Gleiser, Nucl. Phys. {\bf B483},
416--428 (1997).

\bibitem{GM}
M. Gleiser and H.-R. M\"uller, Dartmouth preprint
no. DART-HEP-97/04, hep-lat/9704005. In press, Phys. Lett. B.

\bibitem{HABIB}
F. J. Alexander and S. Habib, Phys. Rev. Lett. {\bf 71},
955 (1993); F. J. Alexander, S. Habib, and A. Kovner, Phys. Rev. E {\bf 48},
4282 (1993); S. Habib, cond-mat/9411058.


\end{thebibliography}
\end{document}